\documentclass[12pt]{amsart}

\usepackage{amssymb}
\usepackage{amsmath,amsthm,amsfonts}  
 \usepackage{graphicx}
\usepackage{pdfsync}
\usepackage{fancyhdr}
\usepackage[applemac]{inputenc}
\usepackage{ulem}
\usepackage{verbatim}
\usepackage{bbm}
\usepackage{float}

\usepackage[default]{frcursive}
\usepackage[T1]{fontenc}

\usepackage{color}
\usepackage{xcolor}
\usepackage[breaklinks,colorlinks,backref]{hyperref}
\hypersetup{
    colorlinks, 
    linktoc=all, 
    linkcolor=red,  
  citecolor=hyptxt,
  urlcolor=blue}
  \hypersetup{
  citebordercolor=,
  filebordercolor=green,
  linkbordercolor=blue
}
\definecolor{hyptxt}{rgb}{0.7, 0.4, 0.9}
\usepackage{bibtopic}

\usepackage[full]{textcomp} 
     \usepackage{fbb} 
     \usepackage[scaled=.95]{cabin}
     \usepackage[varqu,varl]{inconsolata}
     \usepackage[libertine,bigdelims,vvarbb]{newtxmath}
     \usepackage[cal=boondoxo]{mathalfa}
     \usepackage[T1]{fontenc}
\useosf

\definecolor{hervecolor}{rgb}{0.8,0,0.7}

\newcommand{\ket}[1]{|\kern.3ex#1\kern.3ex\rangle}
\newcommand{\bra}[1]{\langle\kern.3ex #1 \kern.3ex|}
\newcommand{\scalar}[2]{\langle\kern.3ex #1 \kern.3ex|\kern.3ex#2\kern.3ex\rangle}

\newcommand{\ii}{\mathsf{i}}

\def\R{\mathbb{R}}

\def\Z {\mathbb{Z}}

\def\lg{\langle }
\def\rg{\rangle }

\def\bu{\mathbbm{1}}

\def\ud{\mathrm{d}}

\def\bi{\widehat{\boldsymbol{\imath}}}
\def\bj{\widehat{\boldsymbol{\jmath}}}

\numberwithin{equation}{section}
\def\mcE{\mathcal{E}}

\newcommand{\sI}{\mathbbm{1}}
\newcommand{\wi}{\widehat{\boldsymbol{\imath}}}
\newcommand{\wj}{\widehat{\boldsymbol{\jmath}}}

\def\bu{\mathbbm{1}}
\def\bi{\widehat{\boldsymbol{\imath}}}
\def\bj{\widehat{\boldsymbol{\jmath}}}

\numberwithin{equation}{section}

\begin{document}
\date{\today}
 
\title[Complexity \& Quantum Orientations]{Quantum circuit complexity for linearly polarised light}
\author[E.~M.F. Curado, S. Faci, J.-P. Gazeau, T. Koide, A. C. Maioli, D. Noguera]{\small 
Evaldo  M.~F. Curado$^{\mathrm{a},\mathrm{b}}$, Sofiane Faci$^{\mathrm{c}}$, Jean-Pierre Gazeau$^{\mathrm{d}}$, Tomoi  Koide$^{\mathrm{e}}$, Alan C. Maioli$^{\mathrm{a}}$, and Diego Noguera$^{\mathrm{a}}$}

\address{\parbox{\linewidth}{\textit{$^{\mathrm{A}}$ Centro Brasileiro de Pesquisas F\'{\i}sicas,}\\
\textit{Rua Xavier Sigaud 150, 22290-180, Rio de Janeiro, RJ, Brazil}}\\}
\address{\parbox{\linewidth}{\textit{$^{\mathrm{B}}$ Instituto Nacional de Ci\^ancia e Tecnologia, Sistemas Complexos,}\\
\textit{Rua Xavier Sigaud 150, 22290-180, Rio de Janeiro, RJ, Brazil}}\\}
\address{\parbox{\linewidth}{\textit{$^{\mathrm{C}}$ Universidade Federal  Fluminense}}\\} 
\address{\parbox{\linewidth}{\textit{$^{\mathrm{D}}$ Universit\'e Paris Cit\'e,}\\ \textit{CNRS, Astroparticule et Cosmologie}  \\
		\textit{75013 Paris, France}}\\}
\address{\parbox{\linewidth}{\textit{$^{\mathrm{E}}$ Instituto de F\'{i}sica, Universidade Federal do Rio de Janeiro, }\\ \textit{ C.P.
68528, 21941-972, Rio de Janeiro, RJ, Brazil}}\\}

\email{\href{mailto:evaldo@cbpf.br}{evaldo@cbpf.br}} 
\email{ \href{mailto:sofiane.faci@gmail.com}{sofiane.faci@gmail.com}} 
\email{ \href{mailto:gazeau@apc.in2p3.fr}{gazeau@apc.in2p3.fr}}
\email{ \href{mailto:tomoikoide@gmail.com}{tomoikoide@gmail.com}}
\email{ \href{mailto:alanmaioli@cbpf.br}{alanmaioli@cbpf.br}}
\email{\href{mailto:diegonoguera.srl@gmail.com}{diegonoguera.srl@gmail.com}}

{\abstract{
In this study, we explore a form of quantum circuit complexity that extends to open systems. To illustrate our methodology, we focus on a basic model where the projective Hilbert space of states is depicted by the set of orientations in the Euclidean plane. Specifically, we investigate the dynamics of mixed quantum states as they undergo interactions with a sequence of gates. Our approach involves the analysis of sequences of real $2\times2$ density matrices. This mathematical model is physically exemplified by the Stokes density matrices, which delineate  the linear polarisation of a quasi-monochromatic light beam, and the gates, which are viewed as quantum polarisers, whose states are also real $2\times2$ density matrices. The interaction between polariser-linearly polarised light is construed within the context of this quantum formalism.    Each density matrix for the light evolves in an approach analogous to a Gorini-Kossakowski-Lindblad-Sudarshan (GKLS) process during the time interval between consecutive gates. Notably, when considering an upper limit for the cost function or tolerance or accuracy, we unearth that the optimal number of gates follows a power-law relationship.
}}

\maketitle

\clearpage

\tableofcontents

\clearpage

\section{Introduction}
\label{intro}

In the realm of computer science, computational complexity (see for instance the inspiring introduction in \cite{jeffermy17}, and references therein) pertains to the smallest number of operations required to accomplish a specific task. 

Quantum computation relies on a sequence of logic gates, each interacting with only a small number of qubits. This sequence defines the evolution operator implemented by the computer. Ideally, this operator is unitary if the system is closed; however, in more realistic scenarios, it may describe the dynamics of an open system.

The difficulty of performing the computation is determined by the number of gates the algorithm utilizes. An algorithm is considered efficient if the number of required gates increases only polynomially with the size of the problem being addressed.

In the closed system case, the process commences by establishing a ``reference'' quantum state denoted as $|\psi_R\rg$, within the framework of some Hilbert space $\mathcal{H}$.  Following this, one proceeds to generate a unitary transformation $U$, which, through its operation, yields the desired ``target'' state, represented as  
\begin{equation}
\label{reftarg}
|\psi_T\rg= U |\psi_R\rg\,. 
\end{equation}
This transformation is essential in achieving the intended outcome.

The unitary transformation $U$ is assembled using a specific set of elementary or universal gates. These gates are applied in sequence to the initial state to tentatively attain the target state. In the context of finite discrete operations this objective is, in general,  not achieved, $|\psi_T\rg \neq  U |\psi_R\rg$.  Consistently, an introduced parameter, known as ``tolerance'' or ``accuracy'',  often marked as $\epsilon$, becomes crucial. This parameter allows for a certain margin of error, acknowledging that achieving exact equality might not always be possible. Instead, the success of the transformation is determined by the degree of proximity between the two states, as measured by a designated distance metric. In other words, the transformation can be considered successful if the two states are sufficiently close, even if they are not precisely equal.
\begin{equation}
\label{disteps}
\Vert  |\psi_T\rg -U |\psi_R\rg \Vert \leq \epsilon\,. 
\end{equation}
Here comes the subtlety of computational complexity. Eq.\,\eqref{disteps} quantifies how hard it is, given one state, to make another, and this cannot always be expressed through the usual inner product of the Hilbert space of states. Even for a single two-state system the computational complexity is rich, see \cite{susskind16,susskind18} for an extensive discussion.
The idea of complexity geometry translates the fact that some controlled operations are harder to perform, such as gates touching many qubits at the same time. Even for a single qubit some elementary universal gates might be harder to implement than others. It is therefore worth studying the simplest possible case.
Naturally, there won't be a singular circuit that accomplishes the desired transformation described in Eq.\;\eqref{reftarg}. In most cases, there will be an abundance of gate sequences that yield the same target state. Nevertheless, the \textit{complexity} of the state $|\psi_T\rg$ can be defined as the \textit{smallest} quantity of gates essential to effect the transformation in Eq.\;\eqref{reftarg} (in fact it is the relative complexity between the reference and target states that matters). This definition encapsulates the number of elementary gates within the most efficient or shortest circuit. Consequently, the challenge lies in discerning this optimal circuit from the myriad of conceivable alternatives.

Nielsen et al \cite{nielsen05,nielsen-etal06,nielsen-dowling07} tackled these questions using a geometric approach inspired by the theory of optimal quantum control and introducing a cost function to evaluate the different feasible paths. Their work draws upon concepts from papers such as those by Gordon and Rice (1997) \cite{gordon-rice97}, Shapiro and Bruner (2003) \cite{shapiro-bruner03}, Rabitz et al (2000) \cite{rabitz-etal00}, and Rice and Zhao (2000) \cite{rice-zhao00}.

Building upon the insights drawn from these concepts, and taking cues from recent studies like those mentioned in \cite{bercugaro19} and \cite{befrigape22}, our work delves into a form of quantum circuit complexity that extends to open systems. It is natural to take into account the influence of the environment, no matter how negligible it might be. In this pursuit, we broaden the formalism to a more encompassing level. Indeed, it is very hard to keep a qubit isolated and in fact dissipate very quickly. Better understanding the dynamics of how this happens and the complexity geometry of open systems is thus fundamental, see for instance \cite{naghiloo20}.

Specifically, we investigate the evolution of mixed quantum states as they undergo interactions with a sequence of gates. Notably, in contrast to the unitary paradigm described in Eq.\;\eqref{reftarg}, our approach deals with a sequence, denoted as $\left(\rho_n\right)_{0\leq n\leq N-1}$, consisting of $N$ density matrices. Each of these matrices evolves within the framework of a Gorini-Kossakowski-Lindblad-Sudarshan (GKLS) process during the time interval between two consecutive gates. To exemplify our approach, we consider a foundational scenario where the Hilbert space of states  is the elementary Euclidean plane
$\R^2$. The corresponding mixed states  $\rho$ are of  the form 
\begin{equation}
\label{rhorp}
\rho_{r,\phi} = \frac{1}{2}\begin{pmatrix}
1 +r  \cos 2\phi    & r\sin 2\phi   \\
   r \sin 2\phi  & 1 -r\cos 2\phi
\end{pmatrix}\, , 
\end{equation}
with $r\in [0,1]\, , \ \phi\in [0,\pi)$.
Pure states correspond to $r=1$, \textit{i.e.}, $
\rho_{1,\phi} =\begin{pmatrix}
  \cos\phi        \\
     \sin\phi   
\end{pmatrix} \begin{pmatrix}
  \cos\phi     &    \sin\phi
  \end{pmatrix}\equiv |\phi\rg\lg \phi|$, 
and $r=0$ is for totally mixed states, \textit{i.e.}, $\rho_{0,\phi} =\bu_2/2$.  
This example offers a significant advantage in that it lends itself to straightforward implementation of both analytical calculations and numerical simulations.  

Hence, we start from a reference state $\rho_{r_R,\phi_R}$ and aim at reaching the target state $\rho_{r_T,\phi_T}$ through a sequence of $N$ gates which modify parameters $r$, $\phi$ and leave the mixed state to interact with its dissipative environment till the next gate. We adopt a von Neumann model of interaction with the gate, as detailed by von Neumann \cite{vonneumann55}, to describe the interaction between linearly polarised light and a quantum polariser. Here, the term ``quantum polariser'' refers to a quantum system represented by real $2\times 2$ density matrices.

In view of comparing various paths in the space of such density matrices, we will use criteria expressed in terms of  euclidean distances between points $(r,\phi)$ which parameterize density matrices.  Moreover, in view of the important outcome of our work, deviating from Nielsen's criteria  will have a negligible impact on our  approach.

In Section \ref{quad}, we expound the application of quantum formalism to the set of orientations (or directions) in the Euclidean plane, treating it as a projective real Hilbert space of quantum states or Dirac kets. Pure and mixed states are  expressed as real $2\times 2$ matrices, each in a direct and intuitive correspondence with points within the upper half-unit disk. The description of quantum observables follows suit, employing symmetric real $2\times 2$ matrices. Subsection \ref{evolution} examines the time evolution of the states discussed earlier. For closed systems, this evolution is governed by simple rotations in the plane. However, when the system interacts with its environment, the time evolution of the density matrix $\rho_{r,\phi}$ is governed by the quantum master equation in the GKLS form, which describes open systems.

In Section \ref{stokes}, we showcase a compelling illustration of pure and mixed states, specifically focusing on real $2\times 2$ matrices that delineate the linear polarisation of light through Stokes parameters (a comprehensive review of classical and quantum polarisations is given in \cite{goldberg21}). This example holds particular significance for our study, infusing a tangible and physical dimension into our exploration of circuit complexity within the confines of the two-dimensional Euclidean framework.

In Section \ref{qumes}, we revisit the above discussion on linear polarisation of light, this time exploring it in the context of the framework of the von Neumann theory of quantum measurement \cite{nello14}. We describe the evolution of the corresponding (Stokes) density matrix as it interacts with a quantum polariser. This evolution is unitary, indicating that there is no collapse of polarisation along a specific direction, as might be expected with classical polarisation.


In Section \ref{sec: polprop}, we offer a brief overview of the physical properties of classical polarisers and clarify how they relate to the quantum framework of light polarization that we introduced in Section \ref{qumes}.

Section \ref{genset} establishes a connection between the previously discussed GKLS equation and the geometric elements such as trace distances between density matrices and the expected straight line  geodesic. These connections are essential for carrying out our program of adapting quantum circuit complexity to the evolution of pure or mixed quantum states in the context of the Euclidean plane.

Section \ref{ansol} provides analytic results enabling us to determine the trajectories derived from the dynamical system presented in the GKLS equation. 

The exploration continues in Section \ref{numap} with concrete numerical investigations. Here, we introduce an algorithmic approach and evaluate its complexity alongside a specific definition for the distance metric. The outcomes stem from the application of multiple polarisers following a distinct strategy, refining the trajectories to align piecewise with desired geodesics. Notably, we observe that our approach effectively adheres to a power law and holds potential applicability across various system types.

Section \ref{conclu} presents a comprehensive discussion regarding the significance and future implications of our results.

Appendices \ref{rpphip},  \ref{polafter} and   \ref{infevol} are dedicated to providing additional technical details that enhance the comprehension of our results.

\section{Quantum formalism for the Euclidean plane}
\label{quad}

The Euclidean plane,  represented as the vector space $\mathbb{R}^2$ with its standard dot product, serves as one of the  straightforward models of a Hilbert space. In this section, we fix the notations and  provide an overview of the real two-dimensional quantum formalism that will be elaborated upon in subsequent sections. Although it may seem rudimentary at first glance, this material adeptly  unveils  a substantial part of subtleties  of the quantum formalism.

 \subsection{Euclidean plane with Dirac notations}

\begin{figure}[H]
\begin{center}
\setlength{\unitlength}{0.1cm} 
\begin{picture}(80,60)
\put(30,20){\vector(1,0){30}} 
\put(30,20){\vector(0,1){30}} 
\put(63, 24){\makebox(0,0){$\wi \equiv |\, \wi\,\rangle$}}
\put(68, 55){\makebox(0,0){$\boldsymbol{v} \equiv |\boldsymbol{v}\rangle \equiv \begin{pmatrix}
      x    \\
      y  
\end{pmatrix}$}}  
\put(30, 16.5){\makebox(0,0){$O$}} 
\put(18, 52){\makebox(0,0){$\wj\equiv | \, \wj\,\rg$}} 
\put(36, 38){\makebox(0,0){$r$}} 
\put(42, 26){\makebox(0,0){$\phi$}}
\put(34,20){\oval(10,15)[tr]}
\thicklines 
\put(30,20){\vector(1,2){18}} 
\end{picture}
\end{center}
\caption{Euclidean plane as a vector space of  Dirac kets. $|\boldsymbol{v}\rg$  denotes  the vector with length $r$ and polar angle $\phi \in [0, 2 \pi)$. For $r=1$ the corresponding unit vector is denoted by $|\phi\rg$.}
\label{euclid}
\end{figure}
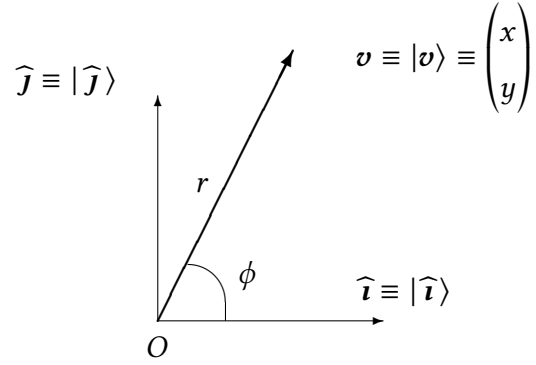

By adopting the Dirac ket notations, as is shown in Fig.\;\ref{euclid}, an orthonormal basis  of the Euclidean plane $\R^2$ is defined by the two vectors 
$| \,\wi\,\rangle$ and $\left| \,\wj  \,\right\rangle$ 
 such that
\begin{equation}
\label{orthframe}
\langle  \wi | \wi  \rangle = 1 =\left\langle \wj \right| \left. \wj \right\rangle\, , \quad \langle \wi\left| \wj \right\rangle = 0\, ,\quad \sI_2 = | \,\wi \, \rangle \langle \, \wi\, |  +  \left| \,\wj\,\right\rangle \left\langle \,\wj\,\right|
\end{equation}

A vector $|\boldsymbol{v}\rg$ with polar angle $\phi \in [0, 2 \pi)$ determines a sense on its support. The latter is a direction or, synonymously, an orientation in the plane. 

\subsection{Pure and mixed  states}
\subsubsection*{Pure states}
In the quantum formalism applied to the Euclidean plane, a unit vector positioned at a polar angle $\phi$ relative to the basis vector $\wi$ symbolizes the pure state $|\phi\rangle$. Its corresponding  orthogonal projector reads as 
 \begin{equation}
\label{projtheta} 
P_{\phi} = | \phi \rangle \langle \phi |  =\mathcal{R}(\phi) |0\rg\lg 0|\mathcal{R}(-\phi) \, . 
\end{equation}
Here,  $\mathcal{R}(\phi)$ denotes the matrix of rotation in the plane by the angle $\phi$,
\begin{equation}
\label{rotmat}
 \mathcal{R}(\phi)= \begin{pmatrix}
  \cos\phi    &  -  \sin\phi  \\
   \sin\phi   &   \cos\phi 
\end{pmatrix}\,.
\end{equation}
$P_{\phi}$ is commonly referred to as a pure state. This  designation is consistent with the fact that the projector ignores sense along the  orientation defined by the polar angle $\phi$,  $P_{\phi}= P_{\phi + \pi}$.  
\subsubsection*{Mixed states}
Mixed states or density matrices $\rho$  are unit trace and  non-negative $2\times 2$ matrices. Thus, they have a spectral decomposition of the form
\begin{equation}
\label{spedec}
\rho = \frac{1+r}{2} |\phi\rg\lg\phi| + \frac{1-r}{2} \left|\phi + \frac{\pi}{2}\right\rg\left\lg\phi+ \frac{\pi}{2}\right|
 \end{equation}
where the   $0\leq r\leq 1$. Hence we can write $\rho$  in terms of the polar coordinates $(r,\phi)$ of a point in the upper half unit disk:
\begin{equation}
\label{standrhomain}
\begin{split}
\rho \equiv\rho_{r,\phi}&= \begin{pmatrix}
  \frac{1}{2}  + \frac{r}{2}\cos2\phi  &   \frac{r}{2}\sin2\phi  \\
\frac{r}{2}\sin2\phi    &   \frac{1}{2}  - \frac{r}{2}\cos2\phi
\end{pmatrix}\\ &= \frac{1}{2}\left(\bu_2 + r \mathcal{R}(\phi)\sigma_3\mathcal{R}(-\phi)\right) \, ,
 \end{split}
\end{equation}
with $\sigma_3= \begin{pmatrix}
  1    &  0  \\
   0   &  -1
\end{pmatrix}$.
The matrix $\rho$ is $\pi$-periodic in variable $\phi$, $\rho_{r,\phi}=\rho_{r,\phi+\pi}$, and we check that for  $r=1$  it is just the orthogonal projector on the unit vector $|\phi\rg$ with polar angle $\phi$, i.e. $ \rho_{1,\phi}= P_{\phi}$.
Also note the covariance property of $\rho_{r,\phi}$ trivially deduced from Eq.\;\eqref{standrhomain}:
\begin{equation}
\label{covrho}
\mathcal{R}(\theta)\rho_{r,\phi}\mathcal{R}(-\theta)= \rho_{r,\phi + \theta}\, . 
\end{equation}
The parameter $r$ can be viewed as a measure of the distance of $\rho$ to the pure state $P_{\phi}$ while $1-r$ measures the degree of ``mixing". A statistical interpretation is made possible through the von Neumann entropy \cite{bengtsson07} defined as
\begin{equation}
\label{VNentr}
S_{\rho}:= -\mathrm{Tr}(\rho\,\ln\rho)=  -\frac{1+r}{2} \ln\frac{1+r}{2} - \frac{1-r}{2} \ln\frac{1-r}{2}\, .
\end{equation}
As a function of $r \in [0,1]$ $S_{\rho}$ is nonnegative, concave  and symmetric with respect to its maximum value $\log 2$ at  $r=0$, \textit{i.e.}, $\rho_0\equiv \bu_2/2$, which describes the state of  completely random orientations, while a pure state $r=1$ yields a vanishing entropy.  

\subsubsection*{Quantum observables}
\label{qobs}
A quantum observable,  $A$ is represented by a $2\times 2$ real symmetric matrix, \textit{i.e.}, a self-adjoint operator. Its spectral decomposition in terms of the two orthogonal projectors associated with its real eigenvalues reads: 
\begin{equation}
\label{spmeasR2}
A= \lambda_\parallel \, P_{\gamma} + \lambda_\bot \, P_{\gamma +\pi/2}\, . 
\end{equation}

Three basic matrices generate the Jordan algebra of all real symmetric $2\times2$ matrices. They are the identity matrix $\bu_2$  and the two real Pauli matrices,
\begin{equation}
\label{Pauli13}
\sigma_1 = \begin{pmatrix}
   0   &  1  \\
    1  &  0
\end{pmatrix}\, ,  \qquad   \sigma_3 = \begin{pmatrix}
   1   & 0   \\
    0  &  -1
\end{pmatrix}\, . 
\end{equation}

\subsection{Time evolution}
\label{evolution}
\subsubsection{Closed systems}
\label{closed}
When is applied to a possibly time-dependent quantum observable $A(t)$ the Heisenberg-Dirac equation reads as 
\begin{equation}
\label{heisenb}
\ii  \frac{\ud}{\ud t} A_H= [A_H, H(t)] +  U(t,t_0)^{\dag}\left(\ii  \,\frac{\partial}{\partial t}A\right)U(t,t_0) \, , 
\end{equation}
where $A_H(t):= U(t,t_0)^{\dag} A(t)U(t,t_0)$. As representing a physical quantity, namely the energy of the system,  the Hamiltonian  $H(t)$ is a self-adjoint operator and generates the unitary evolution operator $\ii \frac{\partial}{\partial t}U(t,t_0) = H(t)U(t,t_0)$.  In the present context  one easily checks \cite{bercugaro19} that the  allowed form of this  Hamiltonian reads:
\begin{equation}
\label{hamsol1}
H(t)= \begin{pmatrix}
   0   & -\ii  \mcE(t)  \\
    \ii  \mcE(t)  &  0
\end{pmatrix} \equiv \ii \mcE(t)\tau_2\, ,
\end{equation}
where the matrix $\tau_2$ is defined in terms of the second Pauli matrix by
\begin{equation}
\label{tau2}
\tau_2= -\ii \sigma_2= \begin{pmatrix}
   0   & -1   \\
   1   &  0
\end{pmatrix}\, .
\end{equation}
As expected,  the corresponding evolution operator acts as a rotation in the plane.
\begin{equation}
\label{evrot}
U(t,t_0) = \exp\ii \int_{t_0}^{t} H(t^{\prime})\ud t^{\prime}= \mathcal{R}\left(\int_{t_0}^{t}\mcE(t^{\prime})\,\ud t^{\prime}\right)\, . 
\end{equation}
Since the Heisenberg-Dirac equation is homogeneous in the imaginary $\ii$, we can  describe  this dynamics in terms of   real numbers only. We just have to deal with  the  real pseudo-Hamiltonian (actually there is no phase space here!):
\begin{equation}
\label{pseudoH}
\widetilde H(t)= \begin{pmatrix}
   0  & -\mcE(t)   \\
   \mcE(t)   &  0
\end{pmatrix} = \mcE(t)  \tau_2\, .
\end{equation}
 One finally obtains the Majorana-like  equation, where only real numbers are involved:
\begin{equation}
\label{pqeveq}
\frac{\ud  A_H}{\ud t} = [A_{H},\widetilde H(t)] + U(t,t_0)^{\dag}\left(\frac{\partial A}{\partial t}\right) U(t,t_0)\, , 
\end{equation}
On the algebraic side, we note that our real quantum dynamics involves the antisymmetric matrix $\tau_2$. Together with $\sigma_1$ and $\sigma_3$ these 3 matrices generate the Lie algebra $\mathfrak{sl}(2,\R)$ of the group SL$(2,\R)$ of $2\times 2$ real matrices with determinant   equal to $1$. 
\begin{equation}
\label{sl2R}
[\sigma_1,\tau_2]= 2\sigma_3\, , \quad [\tau_2,\sigma_3]= 2\sigma_1\, , \quad [\sigma_3,\sigma_1]= -2\tau_2\,. 
\end{equation}

\subsubsection{GKLS Open systems for $\R^2$}
\label{open}
Restricting the form of ``acceptable'' Hamiltonians to Eq.\;\eqref{hamsol1} or Eq.\;\eqref{pseudoH}  is not tenable  since this family does not include the case of quantum \textit{open} systems, particularly when the latter is submitted to measurement. Hence we consider an extension of  Eq.\;\eqref{pqeveq} to a type of \textit{quantum master}   equation, namely the Gorini-Kossakowski-Lindblad-Sudarshan (GKLS) equation \cite{chruscinski-pascazio17}, often just named Lindblad equation \cite{breupetr02},  describing the time evolution of the density matrix $\rho_{r,\phi}$ in Eq.\;\eqref{standrhomain} for an open system.  When applied to the evolution of a quantum observable $A$, the GKLS equation in its diagonal form (which does not restrict its validity)  is expressed as 
\begin{equation}
\label{lindbladA}
\frac{\ud A}{\ud t}=\ii\,[H,A]+\sum _{k}h_k\left[L_{k}AL_{k}^{\dagger }-\frac{1}{2}\left(AL_{k}^{\dagger}L_{k}+L_{k}^{\dagger}L_{k}A\right)\right]\equiv \mathcal{L}(A)\, ,
\end{equation}
where the  $L_k$'s together with the identity form an arbitrary basis of operators and the coefficients $h_k$ are non-negative constants. 
In the present  framework  with no imaginary $\ii$
the three matrices $\sigma_1$, $\tau_2$, $\sigma_3$, together with the identity form a basis for the vector space of  real $2\times 2$ matrices; then the above equation, when  applied to $\rho_{r,\phi}$, becomes
 \begin{equation}
\label{lindbladrho}
\begin{split}
\frac{\ud \rho_{r,\phi}}{\ud t}&=\,[\rho_{r,\phi}, \widetilde{H}]+ h_1 \left(\sigma_1\rho_{r,\phi}\sigma_1- \rho_{r,\phi}\right) + h_2 \left(-\tau_2\rho_{r,\phi}\tau_2 +\rho_{r,\phi}\right) + h_3 \left(\sigma_3\rho_{r,\phi}\sigma_3- \rho_{r,\phi}\right)\\
& = r\left(\frac{\mathcal{E}(t)}{\hbar}\,\sin2\phi - h_1\,\cos2\phi\right)\,\sigma_3 + r\left(-\frac{\mathcal{E}(t)}{\hbar}\,\cos2\phi - h_3\,\sin2\phi\right)\,\sigma_1 + h_2\,\bu_2\,. 
\end{split}
\end{equation}
Since $\dfrac{\ud \rho_{r,\phi}}{\ud t}\equiv \dot  \rho_{r,\phi}= \frac{1}{2}(\dot r \cos2\phi -2r \sin2\phi \,\dot\phi) \sigma_3 + \frac{1}{2}(\dot r \sin2\phi +2r \cos2\phi\,\dot\phi) \sigma_1$, we infer that  $h_2=0$, and, by identification,   one reaches  the first-order differential (dynamical) system
\begin{align}
\label{dynsys1}
  \dot \phi & =\alpha \sin4\phi -\mathcal{E}(t)\, ,  \quad \alpha:= \frac{h_1-h_3}{2} \in \R\, ,\\
 \label{dynsys2} \frac{\dot r}{r} &= -2\alpha \,\cos4\phi - \beta\, ,  \quad \beta:=  h_1+ h_3 > 0\, .
\end{align}  
 Eq.\;\;\eqref{dynsys1} leads to a kind of Ricatti equation  \cite{ince26}. Having in hand a solution $\phi(t)$,  Eq.\;\eqref{dynsys2} is easily integrated and we find
\begin{equation}
\label{sol}
r(t)= r_0 \exp\left[- \beta(t-t_0) -2\alpha \int_{t_0}^{t}\cos4\phi(t^{\prime})\, \ud t^{\prime} \right]\, . 
\end{equation}
Due to the positiveness of $\beta$, we  infer that $r\to 0$, i.e.,  $\rho_{r,\phi}\to \bu_2/2$ as $t\to \infty$, which means that the von Neumann entropy of the open system tends to its maximum at large time, as we can expect.  Note that for $\alpha=0$ the angle $\phi$ behaves 
like for a closed system, 
\begin{equation}
\label{h1eqh3a}
\phi(t)-\phi(t_0)= -\int_{t_0}^{t}\ud t^{\prime}\, \mathcal{E}(t^{\prime})\, , 
\end{equation}
whereas the behaviour of $r$ simplifies to $r(t)= r_0 e^{- 2h_1 (t-t_0)}$.

\section{Example of mixed states: linear polarisation of light and Stokes parameters}
\label{stokes}

A well-established physical interpretation of $2\times2$ real or complex density matrices is found in the context of the Stokes parameters, which characterize the polarization of light. This interpretation is thoroughly explored in \cite{McMaster54} and related references. For a detailed discussion on the significance and measurement of these parameters, see \cite{schaefer07}.

\begin{figure}[H]
\begin{center}
\setlength{\unitlength}{0.1cm} 
\begin{picture}(60,60)
\put(10,10){\line(1,0){50}}
\put(10,10){\line(-1,0){30}}
\put(10,10){\vector(1,0){15}}
\put(10,10){\vector(0,1){15}}
\put(25, 7){\makebox(0,0){$\widehat{\boldsymbol{k}}$}} 
\put(8, 25){\makebox(0,0){$\widehat{\boldsymbol{\jmath}}$}} 
\put(22, 18){\makebox(0,0){$\widehat{\boldsymbol{\imath}}$}} 
\put(10,10){\line(0,1){40}} 
\put(10,10){\line(0,-1){10}} 
\put(10,10){\line(1,1){30}}
\put(10,10){\line(-1,-1){10}} 
\put(10,10){\vector(1,1){12}} 
\put(10,10){\makebox(0,0){$\bullet$}}
\put(10,10){\vector(-2,1){25}} 
\put(-12,26.5){\makebox(0,0){$\mathrm{Re}\left(\overrightarrow{\mathsf{E}}\right)$}}
\put(07, 50){\makebox(0,0){$y$}} 
\put(60, 6.5){\makebox(0,0){$z$}} 
\put(40, 37){\makebox(0,0){$x$}} 
\thicklines 
\end{picture}
\end{center}
\caption{A quasi-monochromatic plane wave propagates to the right along the $z$-axis. The real part of the electric field  lies in the $x-y$ plane.}
\label{proplight}
\end{figure}
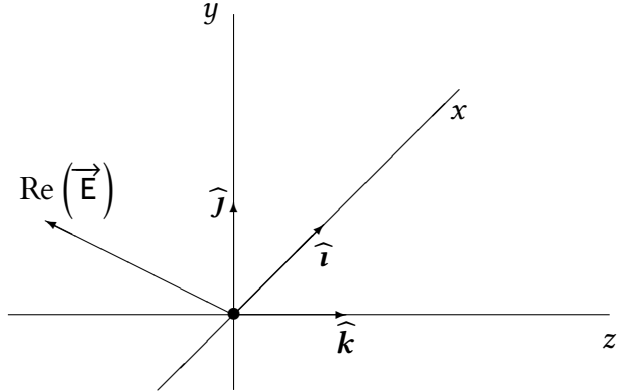

Drawing inspiration from Landau and Lifshitz \cite{Landau75}, we adopt a polarisation tensorial notation $\rho_{\alpha \beta}$ to characterize quasi-monochromatic plane waves propagating along the $z$-axis.  
This framework is expressed in terms of the three Pauli matrices $\sigma_1,\sigma_2, \sigma_3$ and the Stokes parameters $\xi_i$.
\begin{equation}
\label{poltensor}
\begin{split}
\left(\rho_{\alpha \beta}\right) &= \begin{pmatrix}
 \rho_{xx}     &  \rho_{xy}    \\
  \rho_{yx}      &   \rho_{yy} 
\end{pmatrix}=\frac{1}{2}\begin{pmatrix}
1+ \xi_3      & \xi_1 -\ii \xi_2   \\
  \xi_1 + \ii \xi_2     &  1 - \xi_3
\end{pmatrix}\\ & = \frac{1}{2}\left(\bu_2+ \sum_{i=1}^3\xi_i \sigma_i\right)\,, 
\end{split}
\end{equation}   
\begin{equation}
\label{stokland}
\begin{split}
&\xi_1 = r\sin2\phi\,,\quad \xi_2 = A\, , \quad \xi_3 =r \cos2\phi\,,\\
& r= \sqrt{\xi_1^2 + \xi_3^2}\, , \quad \phi= \frac{1}{2}\arctan\frac{\xi_1}{\xi_3}
\end{split}
\end{equation}
The labels $\alpha$, $\beta$ run over the coordinates $(x,y)$ in the plane orthogonal to the $z$-axis. In Eq.\;\eqref{stokland} the parameter $0\leq r\leq 1$ characterizes the degree of maximal linear polarisation whilst $-1\leq A\leq 1$ characterizes the circular polarisation of the beam. Hence, with the notations of Eq.\;\eqref{standrhomain}, the polarisation tensor  Eq.\;\eqref{poltensor} can be written as:
\begin{equation}
\label{rhopol}
\left(\rho_{\alpha \beta}\right) = \rho_{r,\phi} + \frac{A}{2}\sigma_2\,. 
\end{equation}
The Stokes parameters are related to the (complex) electric field component of the light ( Fig.\;\ref{proplight}).  
\begin{equation}
\label{celfield}
\overrightarrow{\mathsf{E}}(t)= \overrightarrow{\mathsf{E}_0}(t)\,e^{\ii \omega t}= \mathsf{E}_x\,\bi + \mathsf{E}_y\,\bj = \left(\mathsf{E}_{\alpha }\right)\,. 
\end{equation}

The time variation of the electric field  $\overrightarrow{\mathsf{E}}(t)$ has average frequency $\omega$, while its $\overrightarrow{\mathsf{E}_0}(t)$ part varies slowly in time. This vector governs the polarisation characteristics of light, which are gauged using Nicol prisms or similar devices. Measurement involves assessing light intensity derived from mean values of quadratic expressions involving field components. Specifically, these expressions, proportional to $\mathsf{E}_{\alpha}\mathsf{E}_{\beta}$, $\mathsf{E}_{\alpha}\mathsf{E}^{\ast}_{\beta}$, and their corresponding complex conjugates, play a crucial role.

Considering $\mathsf{E}_{\alpha}\mathsf{E}_{\beta}=\mathsf{E}_{0\alpha}\mathsf{E}_{0\beta} e^{2\ii \omega t}$ and $\mathsf{E}^{\ast}_{\alpha}\mathsf{E}^{\ast}_{\beta}=\mathsf{E}^{\ast}_{0\alpha}\mathsf{E}^{\ast}_{0\beta} e^{-2\ii \omega t}$, it is evident that these expressions feature rapidly oscillating components. Consequently, their temporal averages $\lg\cdot\rg_t$ equate to zero.

As a result, the complete characterization of partially polarized light can be succinctly conveyed through the description provided by the tensor components:
\begin{equation}
\label{Jalbet}
J_{\alpha\beta}:=\left\lg \mathsf{E}_{0\alpha}\mathsf{E}^{\ast}_{0\beta}\right\rg_t\,.
\end{equation}
The quantity $J=\sum_{\alpha}J_{\alpha\alpha}= \left\lg \vert \mathsf{E}_{0x}\vert^2\right\rg_t + \left\lg \vert \mathsf{E}_{0y}\vert^2\right\rg_t$ determines the intensity of the wave, obtained from the measurement of the energy flux transported by the wave. Since this quantity does not concern the properties of polarisation of the wave, one would rather to deal with the normalized Stokes tensor Eq.\;\eqref{poltensor}, \textit{i.e.}, 
\begin{equation}
\label{Jrho}
\rho_{\alpha \beta} = \frac{J_{\alpha\beta}}{J}\,.
\end{equation}
The light is said completely polarized when the complex amplitude $\overrightarrow{\mathsf{E}}_0$ is time-independent, and so is equal to its time average. Then, the polarisation tensor factorizes as
\begin{equation}
\label{totpol}
\left(\rho_{\alpha \beta}\right) = \frac{1}{J}\begin{pmatrix}
 \vert \mathsf{E}_{0x} \vert^2    & \mathsf{E}_{0x}\mathsf{E}^{\ast}_{0y}   \\
  \mathsf{E}^{\ast}_{0x}\mathsf{E}_{0y}    &   \vert \mathsf{E}_{0y} \vert^2
\end{pmatrix}= \begin{pmatrix}
    \mathsf{E}_{0x} /\sqrt{J}      \\
    \mathsf{E}_{0y}  /\sqrt{J}    
\end{pmatrix}\begin{pmatrix}
  \mathsf{E}^{\ast}_{0x} /\sqrt{J}     & \mathsf{E}^{\ast}_{0y} /\sqrt{J}    \\  
\end{pmatrix} \,,
\end{equation}   
 \textit{i.e.}, is the orthogonal projector along $\overrightarrow{\mathsf{E}}_0$, and, in quantum terms, a pure state.  This case should be put in regard to the other extreme case, namely the non-polarized or natural light, for which all directions in the $x-y$ plane are equivalent:
 \begin{equation}
\label{non-pol}
\left(\rho_{\alpha \beta}\right) = \frac{1}{2}\,\delta_{\alpha\beta}\,.
\end{equation}
In the general case and with the $\xi_i$ coordinates introduced in Eq.\;\eqref{poltensor},  $$\det\left(\rho_{\alpha \beta}\right) = \frac{1}{4}\left(1-\xi_1^2-\xi_2^2 -\xi_3^2\right)\equiv  \frac{1}{4}\left(1-P^2\right)\, , $$ where $0\leq P\leq 1$ is called  the degree of polarisation, from $P=0$ (random polarisation) to $P=1$ (total polarisation). Another extreme case holds  with circular polarisation. Then $\overrightarrow{\mathsf{E}_0}$ is constant and $\mathsf{E}_{0y}= \pm \ii \mathsf{E}_{0x}$, which gives $\left(\rho_{\alpha \beta}\right)= (1/2)(\bu_2\pm\ii \sigma_2)=(1/2)(\bu_2\mp \tau_2)$. As a consequence, the parameter $-1\leq A\leq 1$ in Eq.\;\eqref{stokland} is interpreted as the degree of circular polarisation, with $A=1$ (resp. $A=-1$) for right (resp. left) circular polarisation, and  $A=0$ for linear polarisation. 

The differentiation between linear and circular polarisations precisely aligns with the decomposition of Eq.\;\eqref{poltensor} into its symmetric and antisymmetric components:\begin{equation}
\label{rolroh}
\left(\rho_{\alpha \beta}\right) = \rho_{r,\phi} + \frac{A}{2}\sigma_2 = \frac{1+r}{2} P_{\phi} +  \frac{1-r}{2} P_{\phi + \pi/2} + \frac{A}{2}\sigma_2 \,,
\end{equation}
with notations of   Eq.\;\eqref{standrhomain}. 
In the present work, we  ignore circular polarisation by setting $A$ to zero. Then the polarisation $P=r$ and that simplifies the polarisation tensor down to our real density matrix.

\section{Polarisation of the light through a von Neumann quantum interaction}
\label{qumes}
\subsection{Outcomes of interaction with quantum polariser}
Let us delve into the interaction between a quantum polariser (in the sense given in the introduction) and a partially linear polarised light - an elementary example of a quantum interaction (and, possibly, measurement as well). Our approach is inspired by the example outlined by Peres in \cite{peres90}, but with a specific perspective on two planes and their tensor product. The first plane is the Hilbert space where the states $\rho^P_{s,\theta}$ of the polariser, treated as an orientation modifier, are involved. Notably, the action of the rotation generator $\tau_2=-\ii\sigma_2$ on these states corresponds to a rotation of $\pi/2$:
\begin{equation}
\label{tau2act}
\tau_2\rho^P_{s,\theta}\tau_2^{-1} = -\tau_2\rho^P_{s,\theta}\tau_2 = \rho^P_{s,\theta +\pi/2}\,.
\end{equation}
The second plane is  the Hilbert space on which act the partially linearized polarisation states $\rho^L_{r,\phi}$ of the plane wave crossing the polariser. 
Since the exact interaction time is  ill-determined, the interaction polariser-light  has a time duration  interval $I_M=(t_M-\eta, t_M +\eta)$ centred at $t_M$. This interaction is described by the  (pseudo-) Hamiltonian operator 
\begin{equation}
\label{intham}
\widetilde{H}_{\mathrm{int}}(t)= \, g^{\eta}_M(t)\tau_2 \otimes A^L\,. 
\end{equation}
Here  $g^{\eta}_M$ has support in $I_M$ and approximates the  Dirac peak as $$\lim_{\eta\to 0}\int_{-\infty}^{+\infty}\ud t \, f(t) \,g^{\eta}_M(t)= f(t_M)\,.$$
The action of the polariser on the light is described by the quantum observable  $A^L$, \textit{i.e.},   a symmetric matrix, acting on the states of the light. It has the  spectral  decomposition
\begin{equation}
\label{decAL}
A^L= \lambda_{\parallel}\,P_{\gamma} + \lambda_{\bot}\,P_{\gamma+ \pi/2}\, ,
\end{equation}
 where the direction defined by $P_{\gamma}$ is  interpreted as the filter's orientation, while total absorption occurs along the direction  $\gamma+ \pi/2$. 

The operator \eqref{intham} is the tensor product of an antisymmetric operator, like the pseudo-Hamiltonian \eqref{pseudoH},   for the detector with an operator for the system which is symmetric (\textit{i.e.} Hamiltonian). Nevertheless,  the operator $U(t,t_0)$ defined  for $t_0< t_M-\eta$ as  
\begin{equation}
\label{nevop}
U(t,t_0)= \exp\left[\int_{t_0}^{t}\ud t^{\prime}\, g^{\eta}_M(t^{\prime})\, \tau_2 \otimes A^L\right] = \exp\left[ G_M^\eta(t)\, \tau_2 \otimes A^L\right]\, , 
\end{equation} 
with $G_M^\eta(t)=\int_{t_0}^{t}\ud t^{\prime} \, \,g^{\eta}_M(t^{\prime})$, is an evolution operator. Hence, as soon as $t>t_M+\eta$, $G_M^\eta(t)=1$. From the general formula involving an orthogonal projector $P$,  
\begin{equation}
\label{expP}
\exp(\theta \tau_2 \otimes P) = \mathcal{R}(\theta) \otimes  P + \bu_2\otimes (\bu_2-P)\, , 
\end{equation}
we get the expression:
\begin{equation}
\label{nevop1t}
 U(t,t_0)= \mathcal{R}\left(G_M^\eta(t)\,\lambda_{\parallel}\right) \otimes  P_{\gamma} + \mathcal{R}\left(G_M^\eta(t)\,\lambda_\bot\right)  \otimes  P_{\gamma +\pi/2}\, .
\end{equation}
For $t_0<  t_M-\eta$ and $t> t_M+\eta$, we finally obtain for the evolution operator:
\begin{equation}
\label{nevop1}
 U(t,t_0)= \mathcal{R}\left(\lambda_{\parallel}\right) \otimes  P_{\gamma} + \mathcal{R}\left(\lambda_\bot\right)  \otimes  P_{\gamma +\pi/2}\, . 
\end{equation}
One easily checks that  $U(t,t_0) U(t,t_0)^{\dag} = U(t,t_0)^{\dag} U(t,t_0)  = \bu_2\otimes \bu_2$. 
\subsection{Light after interaction}
After having prepared the polariser in the state $\rho^P_{s_0,\theta_0}$ and its associated observable $A_L$ acting on the light states,  the  action of the operator \eqref{nevop1} on the initial state $\rho^P_{s_0,\theta_0} \otimes \rho^L_{r_0,\phi_0}$ reads for $t> t_M+\eta$:
\begin{equation}
\label{actinUtt0}
\begin{split}
 &U(t,t_0)\,\rho^P_{s_0,\theta_0} \otimes \rho^L_{r_0,\phi_0}\,U(t,t_0)^{\dag}\\ &= \rho^P_{s_0,\theta_0+\lambda_{\parallel}} \otimes \frac{1+r_0\cos2(\gamma-\phi_0)}{2}\,P_{\gamma}\\
 & + \rho^P_{s_0,\theta_0+\lambda_{\bot}} \otimes \frac{1-r_0\cos2(\gamma-\phi_0)}{2}\,P_{\gamma+\pi/2}\\
  &+ \frac{1}{4}\left(\mathcal{R}\left(\lambda_\parallel-\lambda_\bot\right) +s_0\sigma_{2\theta_0 +1}\right)\otimes r_0\sin2(\gamma-\phi_0)\,P_{\gamma}\tau_2\\
 &- \frac{1}{4}\left(\mathcal{R}\left(\lambda_\bot-\lambda_\parallel\right) +s_0\sigma_{2\theta_0 +1}\right)\otimes r_0\sin2(\gamma-\phi_0)\,\tau_2 P_{\gamma}\, . 
 \end{split}
\end{equation}
As expected from the standard  theory of quantum measurement,  this formula indicates that the probability for the observable $A^L$ to orientate along  $\gamma$ is given by the Born rule:
\begin{equation}
\label{pointerphi}
\begin{split}
&\mathrm{Tr}\left[\left(U(t,t_0)\,\rho^P_{s_0,\theta_0} \otimes \rho^L_{r_0,\phi_0}\,U(t,t_0)^{\dag}\right)\left(\bu\otimes P_\gamma\right)\right] \\& =\mathrm{Tr}\left[\rho^L_{r_0,\phi_0} P_\gamma\right]=\frac{1+r_0\cos2(\gamma-\phi_0)}{2}= \frac{1-r_0}{2} + r_0\cos^2 (\gamma-\phi_0)\,,
\end{split}
\end{equation} 
  whereas its probability  to orientate  along   $\gamma + \pi/2$ is:
\begin{equation}
\label{pointerphiorth}
\begin{split}
&\mathrm{Tr}\left[\left(U(t,t_0)\,\rho^P_{s_0,\theta_0} \otimes \rho^L_{r_0,\phi_0}\,U(t,t_0)^{\dag}\right) \left(\bu\otimes P_{\gamma+\pi/2}\right)\right]\\
&= \mathrm{Tr}\left[\rho^L_{r_0,\phi_0} P_{\gamma+\pi/2}\right]= \frac{1-r_0\cos2(\gamma-\phi_0)}{2}= \frac{1-r_0}{2} + r_0\sin^2 (\gamma-\phi_0)\,,
\end{split}
\end{equation} 
 For the completely linear polarisation of the light, i.e. $r_0=1$, we recover the familiar Malus laws, $\cos^2 (\gamma-\phi_0)$ and $\sin^2 (\gamma-\phi_0)$ respectively.
Hence, after the interaction between polariser-partially linear polarised light, the polarisation of the light is described by the new density matrix, obtained by tracing out the polariser part in Eq.\;\eqref{actinUtt0},
\begin{equation}
\label{lpollight}
\rho^L_{r^{\prime},\phi^{\prime}}=\mathrm{Tr}_M\left[U(t,t_0)\,\rho^P_{s_0,\theta_0} \otimes \rho^L_{r_0,\phi_0}\,U(t,t_0)^{\dag}\right]= \frac{1+r^{\prime}}{2} \, P_{\phi^{\prime}} + \frac{1-r^{\prime}}{2} \, P_{\phi^{\prime} + \pi/2}\, . 
\end{equation}
Details of calculations of the parameters $r^{\prime}$ and $\phi^{\prime}$ are given  in Appendix \ref{rpphip}. We obtain:
\begin{align}
\label{rpp}
 & r^{\prime}  =r_0\sqrt{a^2+b^2} = r_0\sqrt{\cos^2 2(\gamma- \phi_0) + \cos^2(\lambda_{\parallel} -  \lambda_{\bot})\sin^2 2(\gamma- \phi_0)}\leq r_0\,, \\
\label{phip} &  \tan \phi^{\prime}  = \frac{\sqrt{a^2 + b^2}-a}{b}\, , \quad b\neq 0 \,, \Leftrightarrow  \tan 2\phi^{\prime}=\frac{b}{a}\, , 
\end{align}
where 
\begin{align}
\label{ap} a&= \cos 2(\gamma- \phi_0)\, \cos2\gamma + \cos(\lambda_{\parallel} -  \lambda_{\bot})\,\sin 2(\gamma- \phi_0)\,\sin2\gamma\,, \\
\label{bp} b&= \cos 2(\gamma- \phi_0)\, \sin2\gamma - \cos(\lambda_{\parallel} -  \lambda_{\bot})\,\sin 2(\gamma- \phi_0)\,\cos2\gamma\,. 
\end{align}
From these expressions, one observes that the initial state $\rho^L_{r_0,\phi_0}$  of the light is not modified if $\gamma=\phi_0$.  
   Although the initial state $\rho^P_{s_0,\theta_0}$ of the polariser serves as an ancilla necessary for implementing the action $A^L$ of the polariser on the light, its own state is altered by the interaction, as detailed in Appendix \ref{polafter}. Additionally, in the Appendix \ref{infevol}, we explore the case of the infinitesimal evolution following the interaction in relation with the GLKS dynamical system.

In order to physically  represent the r\^ole of the polariser as forcing the light to be partially polarised along its  eigendirection  $\gamma$, the angle $\phi^\prime$ of the light state after the action of the polariser  must be the same as  $\gamma$. From this constraint and from Eq.\;\eqref{phip}  we obtain 
\begin{equation}
\label{phigam}
    \tan 2\phi^{\prime}  = \frac{\tan 2 \gamma - \cos(\lambda_{\parallel} -  \lambda_{\bot})\tan 2(\gamma-\phi_0 )}{1+\cos(\lambda_{\parallel} -  \lambda_{\bot})\tan 2(\gamma-\phi_0 )\tan 2 \gamma}= \tan 2 \gamma\,,
\end{equation}
which holds if $\lambda_{\parallel} -  \lambda_{\bot}= (2k+1)\frac{\pi}{2}$. 
It follows for the parameter $ r^\prime$ the attenuation: 
\begin{equation}
\label{r0rp}
    r^\prime=r_0\sqrt{\cos^2 2(\gamma- \phi_0) + \cos^2(\lambda_{\parallel} -  \lambda_{\bot})\sin^2 2(\gamma- \phi_0)} =r_0 \left\vert \cos 2(\gamma-\phi_0)\right\vert\, ,
\end{equation}
which gives the Malus law:
\begin{equation}
\label{malaw2}
\frac{1+  r^\prime}{2}= \frac{1- r_0}{2} + r_0\cos^2(\gamma-\phi_0)\quad \mbox{for} \quad \gamma \leq \phi_0 + \frac{\pi}{2}\, .
\end{equation}

\section{Classical polarisation of the light versus quantum formalism}
\label{sec: polprop}


In the previous section, we discussed the interaction between what we refer to as a quantum polarizer and light, treating it as a unitary process. This approach allows us to calculate the probability of observing each eigenvalue or eigendirection, as expressed in Eq.\;\eqref{decAL}, while deliberately ignoring the inherent irreversibility associated with any realistic quantum measurement - namely, the reduction to an eigenstate of the observable being measured. During such a classical process, it is expected that the polariser enforces one of its eigendirections from Eq.\;\eqref{decAL} onto the light, say $\gamma$, as previously discussed. Consequently, we should explore the possibility of a collapse where $\phi^{\prime} = \gamma$ in Eq.\;\eqref{phip}, either through a projection-valued measure or, in a more sophisticated analysis, through a positive-operator-valued measure. We will now delve deeper into what is generally understood by classical polarization and explore the potential connections with the content presented in Section \ref{qumes}.

A polarisation element is an essential optical device  used to precisely control and manipulate the polarisation state of light \cite{chipman2018polarized}. The primary function of a polarisation element is to transform light between different polarisation states. Polarisers and retarders are the most commonly used and effective polarisation elements in the field of optics. In this context, polarisation properties can be classified into three groups, which are depolarisation, diattenuation, and retardance. Depolarisation can be described as a random reduction in the degree of polarisation, while diattenuation is an alteration in amplitude that depends on polarisation.  Retardance is a modification of the phase related to a specific polarisation. In our work, we  focus on the diattenuation effect to characterize our real polariser and  establish the connection between our quantum  approach to polarisation with these (classical) polarisation properties.  We recall that a polariser is an optical element that transmits light into a desired polarisation state, independent of the incident state. An important aspect of the polariser is that the transmitted orthogonal polarised state is approximately zero for the real polariser and exactly zero for the ideal polariser. Thus, we can define the following ratio for the diattenuation present in a polariser
\begin{equation}\label{eq: diamet}
	D=\frac{T_{max}-T_{min}}{T_{max}+T_{min}}\,,
\end{equation}
where $T_{max}$ is the maximal transmittance when compared among all polarisation states, and $T_{min}$ is the minimum transmittance. The ideal polariser has $D=1$.  On the other hand $D=0$ represents a device without interaction or one that transmits all the polarisation states in the same way. One can recognize the transmittance as the result presented at Eq.\;\eqref{lpollight}, and the respective maximum and minimal values are
\begin{eqnarray}
    T_{max}= \frac{1+r^{\prime}}{2} \ , & \phi^{\prime} = \gamma\,, \\
    T_{min}=\frac{1-r^{\prime}}{2} \ , & \phi^{\prime}= \gamma +\frac{\pi}{2} \, .
\end{eqnarray}
 Therefore, in this context, due to  the normalization of the states by the intensity of the light as described in Eq.\;\eqref{totpol}, the diattenuation $D$ is equivalent to  the degree of depolarisation $r^{\prime}$. 
 It results that the so-called  extinction ratio is given in our case by:
\begin{equation}
    \epsilon_{r}=\frac{1+D}{1-D}=\frac{1+r^{\prime}}{1-r^{\prime}}\,.
\end{equation}

\section{General setting for quantum circuits in the plane}
\label{genset}
\subsection{GKLS equations}
The purpose of the present work  is to adapt the study of complexity of quantum circuit to the material described in the previous sections. Firstly, we aim to deal with quantum states $\rho_{r,\phi}$ and not just pure states, and, secondly, to consider such states as standing for open systems, which means that their parameters $r$ and $\phi$ evolve according to the dynamical system described by Eqs.\;\eqref{dynsys1} and \eqref{dynsys2}. We exclude the case $\beta=0$ since it implies $\alpha= 0$ and just describes the unitary evolution $\phi(t)= \phi_0\exp\left[-\int_{t_0}^t \ud t^{\prime}\,\mathcal{E}(t^{\prime})\right]$ .  

In the scope of this paper, we also deviate from the conventional GKLS equations by broadening their application to encompass evolution models  with time-dependent coefficients.
As regards of this extension   and on a more formal level  \cite{chruscinski-pascazio17,rivas12,alilend07},  one describes the evolution of density operators in the space $\mathcal{B}(\mathcal{H})$  of bounded operators on the Hilbert space $\mathcal{H}$ in terms  of completely positive trace-preserving (CPTP) (dynamical) maps:
\begin{equation}
\label{CPTP}
 \Lambda_{t,t_0} : \mathcal{B}(\mathcal{H})\mapsto \mathcal{B}(\mathcal{H})\, , \quad  \Lambda_{t,t_0}\rho_0= \mathrm{Tr}_{\mathrm{E}}\left(U(t,t_0)\rho_0\otimes\rho_{\mathrm{E}}U^{\dag}(t,t_0) \right)\,, 
\end{equation}
where  $\Lambda_{t,t_0}$ obeys the  time-local but inhomogeneous master equation $\partial_t \Lambda_{t,t_0}= \mathcal{L}_t\circ \Lambda_{t,t_0}$. The density operator $\rho_0$ represents the system's state at the initial time $t_0$, while $\rho_E$ corresponds to the state of the environment. This $\Lambda_{t,t_0}$ is CPTP iff $\mathcal{L}_t$ is of the GLKS form for all $t$, \textit{i.e.}, is  inhomogeneous generalization of a semigroup evolution. In our  case  this means that the parameters $\alpha$ and $\beta$ are time-dependent. Hence we will also deal with  the time-inhomogeneous Lindblad system: 
\begin{align}
\label{dynsyst1}
   \dot \phi &=\alpha(t) \sin4\phi -\mathcal{E}(t)\, ,   \\
 \label{dynsyst2} \frac{\dot r}{r} &= -2\alpha(t) \,\cos4\phi - \beta(t)\, . 
\end{align}    
Like for Eq.\;\eqref{sol}, Eq.\;\eqref{dynsyst2} is easily integrable once given a solution $\phi(t)$  to Eq.\;\eqref{dynsyst1}:
\begin{equation}
\label{ilsol}
r(t)= r_0\exp\left[-\int_{t_0}^t \left(2\alpha(t^{\prime})\,\cos 4\phi(t^{\prime}) + \beta(t^{\prime})\right)\,\ud t^{\prime}\right]\,. 
\end{equation}

\subsection{Metric structure of mixed states for the Euclidean plane}

Concerning the family $\left\{\rho_{r,\phi}\, , \, 0\leq r \leq 1\, , \, \phi\in [0,\pi)\right\}$ of density matrices we are dealing with, it is important for the sequel to describe the metric structure to which they pertain. 

Any density operator $\rho$  has unit trace  norm, 
$\Vert \rho\Vert = \mathrm{tr} \sqrt{\rho \rho^{\dag}} = \mathrm{tr}\sqrt{\rho^2} =1$. 
This property is essential in defining the \textit{trace distance} between quantum states: 
\begin{equation}\label{eq: TraceDistDefinition}
d_{\mathrm{TN}}(\rho_{r,\phi},\rho_{r^\prime,\phi^\prime })=\frac{1}{2} \mathrm{tr}\sqrt{(\rho_{r,\phi}-\rho_{r^\prime,\phi^\prime })^\dagger(\rho_{r,\phi}-\rho_{r^\prime,\phi^\prime })}=\frac{1}{2}\sum_{i=1}^2 \vert \lambda_i \vert \, ,
\end{equation}
where $\lambda_i$ is the i-th eigenvalue of the the matrix $(\rho_{r,\phi}-\rho_{r^\prime,\phi^\prime })$.  It is easy to show that
\begin{equation}
\label{dTN}
    d_{\mathrm{TN}}(\rho_{r,\phi},\rho_{r^\prime,\phi^\prime })=\frac{1}{2} \sqrt{r^2 + {r^{\prime}}^2 -2 r r^{\prime}\cos(2\phi-2\phi^{\prime})}= \frac{1}{2}\Vert (r,2\phi) - (r^{\prime},2\phi^{\prime})\Vert\, , 
\end{equation}
i.e., is one-half of the Euclidean distance between the points $(r,2\phi)$ and $(r^{\prime},2\phi^{\prime})$. Note that for $r=r^{\prime}$  this expression simplifies to 
\begin{equation}
\label{dTNrr}
    d_{\mathrm{TN}}(\rho_{r,\phi},\rho_{r,\phi^\prime })=r \vert \sin(\phi-\phi^{\prime})\vert\,. 
    \end{equation}
Geodesic curves corresponding to this metric are found as the  solution of the the Euler-Lagrange equations:
\begin{equation}\label{eq: geodesicsrelations1b}
    \ddot \phi(t)+2 \frac{\dot r(t)}{r(t)}\dot\phi (t)=0\,,  \quad \frac{\ddot r (t)}{r(t)}- \dot \phi^2(t)=0\, .
\end{equation}
As one might expect, they are straight lines  
and are described in polar coordinates  by: 
\begin{equation}\label{eq: geodesicsTraceNorm}
 r=\frac{1}{C_3 \cos  \phi +C_4 \sin  \phi}\, ,
\end{equation}
where the coefficients can be obtained via the chosen reference and target states
\begin{equation}
    C_3 = \frac{r_T \sin \phi_T-r_R \sin \phi_R}{r_Tr_R\sin(\phi_T-\phi_R)}\,, \quad 
    C_4 = \frac{-r_T \cos \phi_T+ r_R \cos \phi_R}{r_Tr_R\sin(\phi_T-\phi_R)}\,,
\end{equation}
 Note that one can rewrite the trace  distance $d_{\mathrm{TN}}$ using the Stokes parameters presented in  Eq.\;\eqref{stokland}
\begin{equation}\label{eq: TraceNormDistanceStokes}
    d_{\mathrm{TN}}(\rho_{r,\phi},\rho_{r^\prime,\phi^\prime })= \frac{1}{2}\sqrt{(\xi_1-\xi_1^\prime)^2+(\xi_3-\xi_3^\prime)^2},
\end{equation}
where $\xi_1^2+\xi_3^2\leq 1$, and $0\leq d_{\mathrm{TN}}\leq 1$.

  \section{Some analytic solutions to GKLS equations}
\label{ansol}

We exhibit some analytic solutions for the dynamical system described by  Eqs.\;\eqref{dynsys1} and \eqref{dynsys2} describing the time-homogeneous case (constant coefficients), and Eqs.\;\eqref{dynsyst1}-\eqref{dynsyst2} for inhomogeneous configuration. We first explore  specific cases of the homogeneous GLKS equations for which the energy $\mathcal{E}$ is constant, and then for which  the the angular velocity $\dot\phi$ is constant. Second we study the time-inhomogeneous case when  $\phi$ is constant.

\subsection{Constant energy $\mathcal{E}$} 
\label{cstpar}
The equations \eqref{dynsys1} and \eqref{dynsys2} have constant coefficients $\alpha$ and $\beta$. Let us consider the energy parameter $\mathcal{E}$ in Eq.\;\eqref{dynsys1} as constant. Integrating the system yields:
\begin{equation} \label{eq: phi integral }
    \int_{\phi(t_0)}^{\phi(t)} \frac{\ud\phi^{\prime}}{\alpha \sin 4\phi^{\prime}- \mathcal{E} }= t-t_0\, ,
\end{equation}
The general form of this integral is found in \cite{gradshteyn2014table},     2.551.3:   
\begin{align}
    \int \frac{\ud x}{a+b \sin x} =& \nonumber \\
\label{altb}    & \frac{2}{\sqrt{a^2-b^2}}\arctan \left[ \frac{a \tan \frac{x}{2}+b}{\sqrt{a^2-b^2}}\right] \quad {\rm if} \ a^2>b^2\,, \\
 \label{astb}   &   \frac{1}{\sqrt{b^2-a^2}} \ln \left[ \frac{a \tan \frac{x}{2}+b-\sqrt{b^2-a^2}}{ a \tan \frac{x}{2}+b+\sqrt{b^2-a^2} } \right] \quad {\rm if} \ a^2<b^2\, .
\end{align}
\subsubsection*{Case $\mathcal{E}^2  >\alpha^2$}
\begin{figure}[htp!]
    \centering
    \includegraphics[width=.49\textwidth]{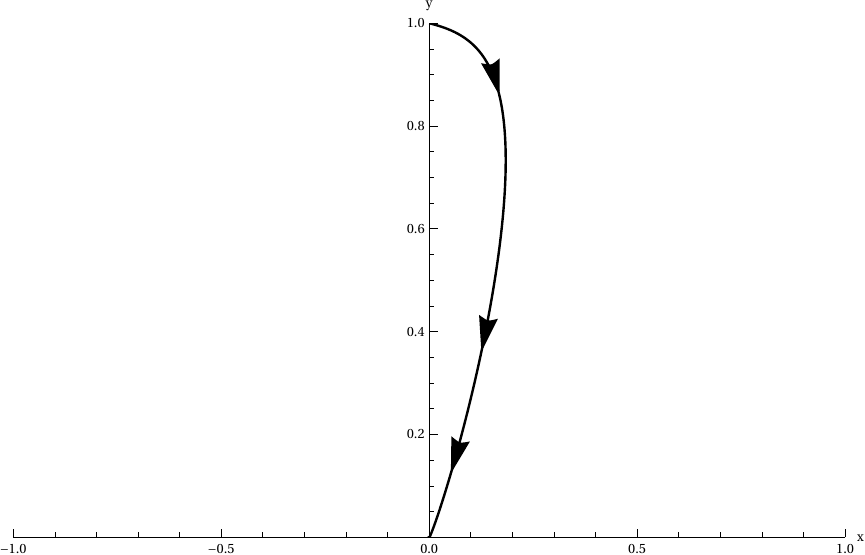}
    \includegraphics[width=.49\textwidth]{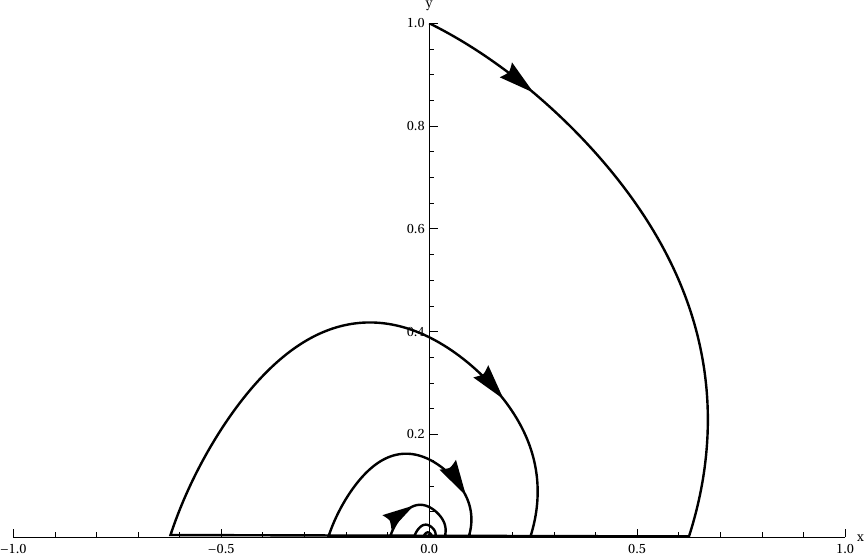}
    \caption{Plot of the trajectory of the density matrix with initial conditions $r(0)=1$ and $\phi(0)=\pi/2$. Both figures have the energy $\mathcal{E}=10$, but different parameters $\alpha=-9$, $\beta=20$ (left), $\alpha=0.5$, and $\beta=3$ (right).}
    \label{fig: TimeEvoLargerE}
\end{figure}

 The application of Eq.\;\eqref{altb} to Eq.\;\eqref{eq: phi integral } gives the solution:
\begin{equation} \label{eq: phiresulta2maiorb2}
    \phi(t)= \frac{1}{2} \arctan \left[ 
\frac{- \omega_1}{\mathcal{E}}\tan (2\omega_1 \Delta t +c_0)+\frac{\alpha}{\mathcal{E}} \right]+ \frac{n \pi}{2}\,, 
\end{equation}
where
\begin{equation}
\Delta t = t -t_0\, , \quad  \omega_1 = \sqrt{\mathcal{E}^2-\alpha^2 }\, , \quad  c_0 = \arctan \left[ \frac{-\mathcal{E} \tan 2 \phi(t_0)+\alpha}{\omega_1} \right]\, . 
\end{equation}
Here $n\in \Z$ must be carefully chosen to insure the continuity of $\phi(t)$ in the interval $ [ 0,\pi )$. In Fig. \ref{fig: TimeEvoLargerE}, we plot the trajectory of the density matrix in the upper half-unit disk, where $\textbf{r}(t)=( r(t) \cos\phi(t),r(t) \sin\phi(t) )$ for two representative cases. The $r(t)$ values were obtained via numerical integration of the $\cos4\phi(t)$ in Eqs. \;\eqref{sol} or \;\eqref{ilsol} for constant $\beta$. In this context, it becomes evident that making the  parameter values vary leads to disparate trajectories, whether exhibiting cyclic patterns or not, across the upper half-unit disk. This connection becomes explicit when one identifies the constant $\omega_1$ with an angular frequency. Then, the period $T$ of the $\phi$ cycle is  $T=\pi/(2 \omega_1 )$, where $\pi$ is the periodicity of the tangent function. 

\subsubsection*{Case $\mathcal{E}^2  < \alpha^2$} 
Eq.\;\eqref{eq: phi integral } now leads to the solution:
\begin{equation}\label{eq: phiresulta2menorb2}
    \phi(t)= \frac{1}{2} \arctan \left[ 
  \frac{\alpha}{\mathcal{E}}+\frac{\gamma}{\mathcal{E}}\coth(2\gamma \Delta t +c_2 )\right]+\frac{n\pi}{2}\,,
\end{equation}
where we have introduced  
\begin{equation}
 \gamma = \sqrt{\alpha^2-\mathcal{E}^2} \, , \quad   c_2=\frac{1}{2}\ln\left[ \frac{-\mathcal{E}\tan2\phi(0)+\alpha-\gamma}{-\mathcal{E}\tan2\phi(0)+\alpha+\gamma}\right]\, .
\end{equation}
The  integer $n$  is responsible for the continuity of $\phi(t)$ in the upper half-unit disk. The parameter $\gamma$ can be viewed as an attenuation constant for the variable $\phi$, given the behavior of the $\coth$ function, which tends towards unity for extended time intervals.

\subsection{Constant angle $\phi$}
\label{cstphi}
\begin{figure}[H]
    \centering
    \includegraphics[width=.7\textwidth]{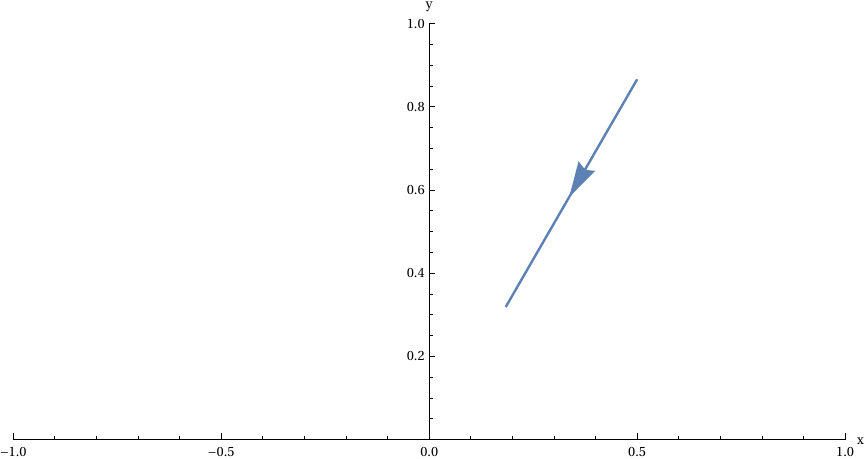}
    \caption{Straight line trajectory that obeys Eq.\;\eqref{eq: straigthlineorigin} with $\phi_R=\pi/3$,  $r_R=1$, $\alpha=1$, and $\beta=2$.}
    \label{fig: straightline}
\end{figure}

Considering  Eqs.\;\eqref{dynsys1}-\eqref{dynsys2} of the inhomogeneous GLKS system, we  analyze the case where $\dot \phi=0$. Then, the $\phi$ variable  does  not change over time, and this  implies
\begin{equation}\label{eq: straigthlineorigin}
    \mathcal{E}(t)= \alpha(t) \sin 4\phi_R\, ,
\end{equation}
where $\phi_R$ is the initial condition for $\phi$. We have now to deal with the inhomogeneous GLKS case. Then the radial function reads:
\begin{equation}
    r(t) = r_R \exp \left[-\int_{0}^t \left(2\alpha(t^{\prime})\,\cos 4\phi_R+\beta(t^{\prime})\right)\ud t^{\prime}\right]\, ,
\end{equation}
where $r_R = r(t_0)$ is the reference state $r$ value. 
This integrand  decreases over time, because $2\alpha(t)\cos 4\phi_R+ \beta(t)=h_1(t)(1 + \cos 4\phi_R) + h_2(t)(1 - \cos 4\phi_R) > 0$ for $h_1(t)$ and $h_2(t)$ being positive.  As a consequence, the trajectory in the upper half disk forms a straight line leading to the origin, as illustrated in Fig. \ref{fig: straightline}, irrespective of whether the parameters are constant or time-dependent. 
A special situation happens when $2\alpha(t)\cos 4\phi_R+ \beta(t)=0$. This leads to $r(t) = r_R$, the circuit trajectory thus collapses to a single fixed point.
\subsection{Constant $\dot \phi$}
\label{cstdotphi}


When $\alpha= 0$ and both $\beta$ and $\mathcal{E}$ are constant, Eqs. \;\eqref{dynsys1} and \;\eqref{dynsys2} become $\dot \phi = -\mathcal{E}$  and $\dot r=-\beta$, respectively. The
two equations are simply solved through $\phi(t)= \phi_R - \mathcal{E}\Delta  t $ and $ r = r_R e^{-\beta\Delta t}$.
Therefore, within the quantum circuit framework, for which $t_R$ and $t_T$ are references and target times respectively, one can directly determine the values of the system parameters  in function of the reference and target states variables:
\begin{equation}\label{eq: ParametersAlphazero}
    \mathcal{E}=\frac{\phi_R - \phi_T}{t_T-t_R} , \qquad \beta=\frac{1}{t_T-t_R}\ln\frac{r_R}{r_T}\, .
\end{equation}

Note that $\phi$ is restricted to the upper half-plane, $\phi \in  [0, \pi)$  and so the solution is $\pi$ cyclic.

%

\section{Numerical approach to complexity circuit}
\label{numap}
We now enter into the heart of our project in dealing with sequences of $N$ gates (\textit{e.g.}, quantum polarisers) 
through the path followed by the state $\rho^L_{r,\phi}$ from the referent $\rho^L_{r_{\mathrm{R}},\phi_{\mathrm{R}}}$ to the target $\rho^L_{r_{\mathrm{T}},\phi_{\mathrm{T}}}$. Our objective is to ensure that the states closely follow the geodesics associated with the trace distance metric \eqref{eq: TraceDistDefinition}, based on the reasonable assumption that these geodesics represent the optimal circuits. Specifically, we aim to approximate the geodesics in Eq.\;\eqref{eq: geodesicsTraceNorm} using a sequence of trajectories, each with constant parameters. This condition arises from the fact that such trajectories are the easiest to implement in dealing with the GLKS dynamics. 

Precisely, the curve initiates from the reference state $\rho^L(r_R,\phi_R)$ and ends at a target state $\rho^L(r_T,\phi_T)$. Assuming that the environment interacts with the state, our system is viewed as an open system whose evolution is determined by the GKLS equation (Subsubsec. \ref{open}). We consider an isotropic environment, where the decay rates are constant $h_1=h_3=1$, i.e., $\alpha=0$ and $\beta= 2$. Moreover we  choose a constant pseudo-Hamiltonian with $\mathcal{E}=-2$. Without the action of a quantum polariser, the system  evolves according to the GKLS equations, and it  naturally changes its  state in the sense of depolarisation. We then impose the action of a sequence of quantum polarisers (gates) in order to force the light state to adhere closely to the geodesic. Both the geodesic and the trajectory with constant $\mathcal{E}, \beta, \alpha$ have $\dot{r}<0$. Along these lines, we know that $r(t)$, as a monotonically decreasing function,  is one-to-one. Then, rather than calculating the distance from the trajectory to the geodesic at each moment, we compare the geodesic state with the time-evolving state at the same radial coordinate $r$. It results the simplified expression for the trace distance Eq.\;\eqref{dTNrr}:
\begin{equation}
\label{trdistrrp}
   d_{\mathrm{geo}}\left(\rho^L_{r,\phi} \,, \rho^L_{r,\phi_{\mathrm{geo}}}\right)=r \left\vert \sin(\phi-\phi_{\mathrm{geo}}) \right\vert\, ,
\end{equation}
where $\phi_{\mathrm{geo}}$ is the angle parameter of the geodesic state.
We want to  identify the minimum number of gates necessary to maintain the constant parameter path within a defined accuracy (or tolerance) threshold $\epsilon$
\begin{equation}\label{eq: accu}
    d_{\mathrm{geo}}\left(\rho^L_{r,\phi} \ ,  \ \rho^L_{r,\phi_{\mathrm{geo}}}\right)\leq \epsilon\, .
\end{equation}

\begin{figure}[H]
    \centering
\includegraphics[width=.99\textwidth]{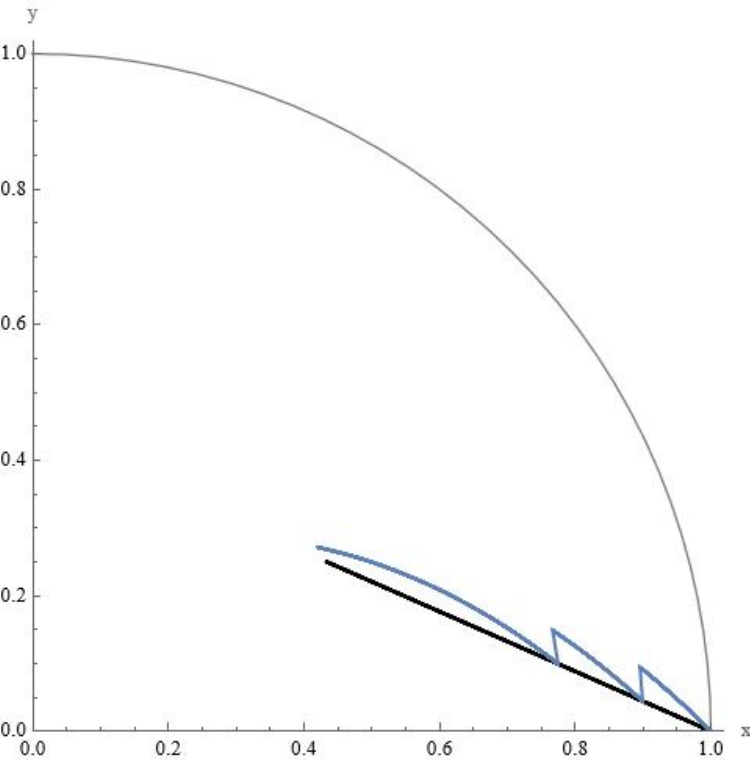}
    \caption{The black line represents the geodesic related to the trace distance and for example (a) of the table \ref{tab: examplevalues}, and the blue curves represent the trajectory of the quantum state evolved accordingly with GLKS equations. When the distance reaches a chosen accuracy $\epsilon=0.05$, we place a quantum polariser to change the angle of the polarisation to be the same as the one at the geodesics. The action of the polarization can be viewed as the abruptly change in the quantum state path.}
    \label{fig: exampleATracedist}
\end{figure}

Whenever the state $\rho^L_{r,\phi}$ deviates from $\rho^L_{r,\phi_{\mathrm{geo}}}$ by more than $\epsilon$, we implement the action of a quantum polariser with an angle $\gamma=\phi_{\mathrm{geo}}$ to bring the state back close to the geodesic. Essentially, this strategy involves applying the quantum polariser's action $A^L$ whenever the state deviates from the geodesic by more than the maximum distance defined in Eq.\;\eqref{eq: accu}. To demonstrate the application of this strategy, we select several representative cases, as defined in Table \ref{tab: examplevalues}, and illustrate them in Fig. \ref{fig: exampleATracedist}.

The values of the parameters of the target and reference states are the ones given  in the example (a) of Table \ref{tab: examplevalues}.  We choose an accuracy of $\epsilon=0.05$. Whenever the trace distance outstretches the accuracy we employ the polariser to put the state at the same angle as the geodesic. 

\begin{table}[H]
    \centering
    \begin{tabular}{|c|c|c|c|c|}
    \hline
       Example  & $\phi_R$ & $r_R$ & $\phi_{T}$ & $r_{T}$  \\ \hline \hline
        a & $0$ & $1$ & $\pi/6$ & $0.5$ \\
        b & $\pi/3$ & $1$ & $\pi/2$ & $0.5$ \\
        c & $\pi/4$ & $1$ & $\pi/12$ & $0.5$  \\
        d & $11\pi/12$ & $1$ & $3\pi/4$ & $0.5$\\
        \hline
    \end{tabular}
    \caption{Our selected values for the reference state variables $\phi_R$, $r_R$ (all are for pure states), and the target state variables $\phi_{T}$, $r_{T}$.}
    \label{tab: examplevalues}
\end{table}

\begin{figure}[H]
    \centering
\includegraphics[width=.99\textwidth]{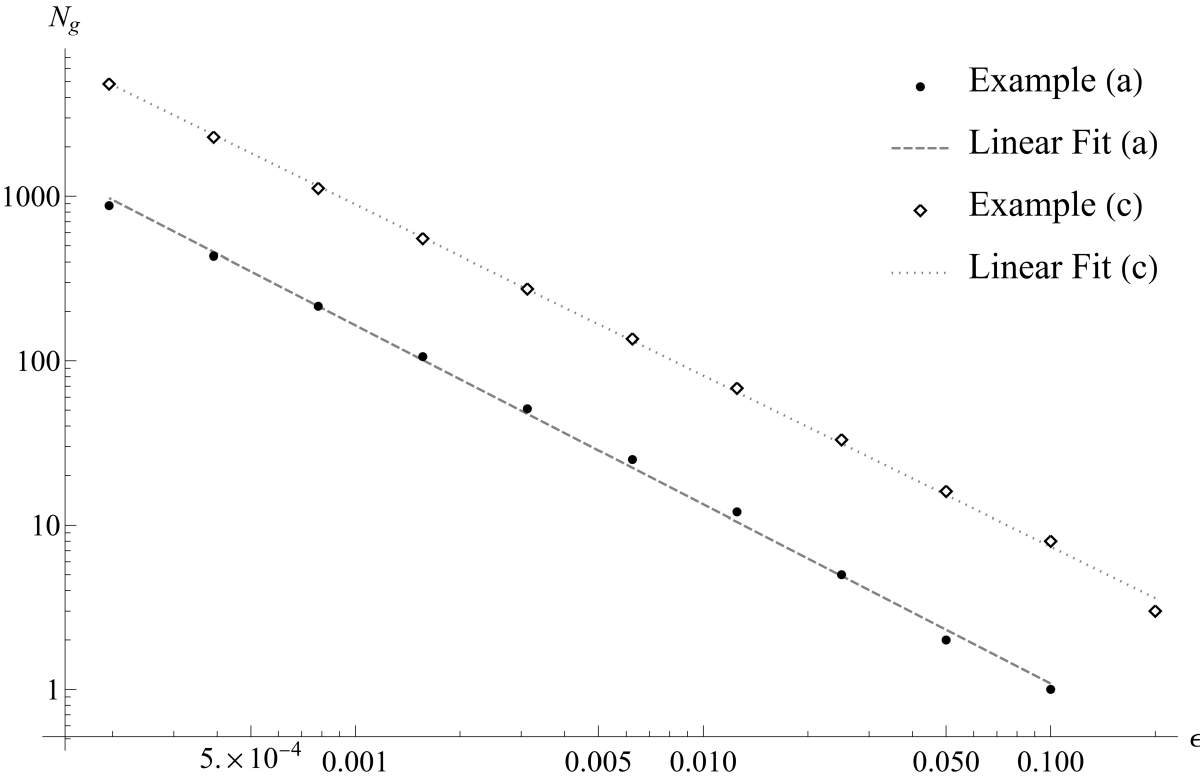}
    \caption{The Log-Log plot of the number of polarisers for each accuracy related to examples (a) and (c) of the table \ref{tab: examplevalues}. The metric here is defined by the trace distance. The pairs of examples (a) and (b), and (c) and (d) have the same data. Therefore, they share the same results for the estimates of the linear fit: (a) $m=-1.05469$, $n=-1.09012$, (c) $m=-0.171026$, $n= -1.04029$. The examples (b) and (d) are omitted in the figure.}
    \label{fig: loglogfixeddistanceTRACE}
\end{figure}

To analyze the complexity of this quantum circuit, we apply the same procedure to all the examples listed in Table \ref{tab: examplevalues}. By varying the accuracy parameter, we determine the corresponding number of gates required. The results are presented in the Log-Log plot in Fig. \ref{fig: loglogfixeddistanceTRACE}, along with a linear fit to the data.

It can be observed that, for examples (a) and (b) in Table \ref{tab: examplevalues}, the same number of gates is needed to keep the path close to the geodesics, regardless of the accuracy parameter. Similarly, examples (c) and (d) exhibit analogous behavior. The distinction between the pairs of examples (a,b) and (c,d) arises from the variation of $\phi$
in the geodesics compared to the constant parameter path. Specifically, in examples (a) and (b), both the geodesics and the trajectories described by the GKLS equations have 
$\dot\phi>0$, whereas in examples (c) and (d), the geodesics have $\dot\phi<0$. 

From the linear fit parameters in $\log N_g=m+n \log \epsilon$, we observe that in both cases (a,b) and (c,d) the circuits follow a power law:
\begin{align}
    N_g &\approx 10^{-1.05} \times \epsilon^{-1.09} \quad   {\rm examples \ (a,b)},  \\
    N_g &\approx 10^{-0.17} \times \epsilon^{-1.04} \quad {\rm examples \ (c,d)}\, . 
\end{align}
These power-law behaviors are consistent with the typical patterns observed in efficient algorithms, avoiding the exponential growth commonly associated with increased complexity. We regard these findings as the most significant insights from our exploration of this simple model of an open quantum circuit.

\section{Conclusion}
\label{conclu}
We have developed a quantum framework for the Euclidean plane, offering a detailed overview of the fundamental elements necessary for its representation. This framework incorporates the use of Dirac notation to describe the Euclidean plane, defines both pure and mixed states, and clarifies the role of quantum observables. We further demonstrate how this formalism can be interpreted through the lens of linear polarisation variation of light interacting with a concept we define as a quantum polariser. Additionally, we explore the time evolution of both closed and open quantum systems, with the latter being governed by the GKLS (Lindblad) equation, which we reformulate as a dynamical system.

In this context, we have presented analytic solutions for the dynamical system resulting from the homogeneous GKLS equation with constant energy and constant angular velocity  $\dot\phi$, and specific cases for the non-homogeneous GKLS equation with constant angle $\phi$.
With the help of those constant parameter solutions, we have investigated the number of gates $N_g$ requested to suitably approximate  the geodesic. In other words, for an open quantum system, where states are represented as points on the upper half of the unit disk,  we addressed the problem of achieving a  target mixed state $\rho_T$ from a reference (generally pure) state $\rho_R$   by attempting to follow as closely as possible,  a geodesic determined by the trace distance metric. Such a geodesic the latter is naturally  considered as an optimal circuit.  

 We have proposed a strategy to determine $N_g$ for a given geometric configuration  and desired accuracy. Notably, we observed that the minimum number of gates $N_g$ follows a power-law scaling with respect to the accuracy, while keeping the maximum allowed distance constant. This behavior aligns with  patterns commonly seen in efficient algorithms, effectively avoiding exponential growth. 
   
 It is crucial to highlight the broad applicability of this strategy, as it can be implemented with alternative definitions of distance. This concept can be extended to closed or open systems featuring one or more qubits, and, of course, systems that are described by quantum mechanics with complex numbers, like spin $1/2$ systems, two level atoms, and beyond. Other potential applications of the present approach include addressing time optimality questions in quantum evolution, as explored in works such as \cite{carlini06,koike22}. Additionally, this approach could contribute to the rapidly advancing field of quantum thermodynamics; relevant examples in this context can be found in \cite{kammerlander16,elouard17}.

\newpage
\appendix
\section{Density matrix for  light after polarisation interaction}
\label{rpphip}
We here give the details of the calculations of the parameters $r^{\prime}$ and $\phi^{\prime}$ in Eq.\;\eqref{lpollight}. With the notations
\begin{align*}
    A^L &= \lambda_{\parallel}\,P_{\gamma} + \lambda_{\bot}\,P_{\gamma+ \pi/2}\,,   \\
   \delta &=   \lambda_{\parallel} -  \lambda_{\bot}\, , \\
   \zeta&= 2(\gamma- \phi_0)\, , 
\end{align*}
and from the expression Eq.\;\eqref{actinUtt0},
we successively have:
\begin{align*}
\rho^L_{r^{\prime},\phi^{\prime}}
&= \mathrm{Tr}_M\left[U(t,t_0)\,\rho^M_{s_0,\theta_0} \otimes \rho^L_{r_0,\phi_0}\,U(t,t_0)^{\dag}\right]\\ \nonumber
&=   \frac{1+r_0\cos\zeta}{2}\,P_{\gamma} +  \frac{1-r_0\cos\zeta}{2}\,P_{\gamma+\pi/2}+ \frac{r_0}{2}\cos\delta\, \sin\zeta\,\left[P_{\gamma},\tau_2\right] \\
&= \frac{\bu}{2} + \frac{r_0}{2}
\begin{pmatrix}
   \cos\zeta   &  \cos\delta\,\sin\zeta  \\
    -\cos\delta\,\sin\zeta  &  \cos\zeta
\end{pmatrix} \,\sigma_{2\gamma}\equiv  
  \frac{\bu}{2} + \frac{r_0}{2}\,\begin{pmatrix}
   a   & b   \\
   b   &  -a
\end{pmatrix}\, , 
\end{align*}
with 
\begin{align*}
a&= \cos\zeta\, \cos2\gamma + \cos\delta\,\sin\zeta\,\sin2\gamma\,, \\
b&= \cos\zeta\, \sin2\gamma - \cos\delta\,\sin\zeta\,\cos2\gamma\,. 
\end{align*}
The spectral decomposition  of the matrix $\begin{pmatrix}
   a   & b   \\
   b   &  -a
\end{pmatrix}$ reads as
\begin{equation*}
\label{ }
\begin{pmatrix}
   a   & b   \\
   b   &  -a
\end{pmatrix} = \sqrt{a^2 +b^2} \left(P_{\varepsilon} - P_{\varepsilon +\pi/2}\right)= \sqrt{a^2 +b^2}\,\sigma_{2\varepsilon}\, , 
\end{equation*} 
with 
\begin{equation*}
\label{ }
\tan\varepsilon= \frac{\sqrt{a^2 + b^2}-a}{b}\, , \quad b\neq 0\, . 
\end{equation*}
Thus, by identification with $\rho^L_{r^{\prime},\phi^{\prime}}= \frac{1+r^{\prime}}{2} \, P_{\phi^{\prime}} + \frac{1-r^{\prime}}{2} \, P_{\phi^{\prime} + \pi/2}$ we find:
\begin{equation*}
\label{ }
r^{\prime}=r_0\sqrt{a^2+b^2} = r_0\sqrt{\cos^2\zeta + \cos^2\delta\sin^2\zeta}\leq r_0\, , \quad \phi^{\prime}=\varepsilon\,. 
\end{equation*}

\section{Quantum polariser after interaction}
\label{polafter}
It is interesting to see what the state of the quantum polariser  becomes after interaction:
\begin{equation}
\label{lpolpol1}
\rho^P_{s^{\prime},\theta^{\prime}}=\mathrm{Tr}_L\left[U(t,t_0)\,\rho^P_{s_0,\theta_0} \otimes \rho^L_{r_0,\phi_0}\,U(t,t_0)^{\dag}\right]= \frac{1+s^{\prime}}{2} \, P_{\theta^{\prime}} + \frac{1-s^{\prime}}{2} \, P_{\theta^{\prime} + \pi/2}\, . 
\end{equation}
First we obtain the convex combination of two states:
\begin{equation}
\label{polpol2}
\rho^P_{s^{\prime},\theta^{\prime}}= \frac{1+r_0\cos2(\gamma-\phi_0)}{2} \, \rho^P_{s_0,\theta_0+\lambda_\parallel} + \frac{1-r_0\cos2(\gamma-\phi_0)}{2} \, \rho^P_{s_0,\theta_0+ \lambda_\bot}\,. 
\end{equation}
Then, we easily obtain $s^{\prime}$ and $\theta^{\prime}$ from the general convex superposition formula: 
\begin{equation}
\label{convcomb}
\begin{split}
&\lambda_1\,\rho_{r_1,\phi_1} + \lambda_2\,\rho_{r_2,\phi_2}= \rho_{r,\phi}\, ,  \quad \lambda_1 + \lambda_2=1\,, \quad \lambda_1\, ,\, \lambda_2 \geq 0\, , \\
&r= \sqrt{\lambda^2_1 r^2_1+ \lambda^2_2 r^2_2 - 2 \lambda_1 \lambda_2r_1r_2 \cos2\left(\phi_1-\phi_2\right)} \\
&e^{2\ii \phi} = \frac{\lambda_1 r_1 e^{2\ii \phi_1} +  \lambda_2 r_2 e^{2\ii \phi_2}}{\sqrt{\lambda^2_1 r^2_1+ \lambda^2_2 r^2_2 - 2 \lambda_1 \lambda_2r_1r_2 \cos2\left(\phi_1-\phi_2\right)}}\,. 
\end{split}
\end{equation}
Applied to the case Eq.\;\eqref{polpol2} we obtain:
\begin{equation}
\label{polpol3}
\begin{split}
s^{\prime}&=\frac{s_0}{\sqrt{2}}\sqrt{\left(1+ r^2_0\cos^22(\gamma-\phi_0) \right) -\left(1- r^2_0\cos^22(\gamma-\phi_0) \right)\cos2\left(\lambda_{\parallel}-\lambda_{\bot}\right)}\, , \\
e^{2\ii \theta^{\prime}}&= \sqrt{2}e^{2\ii \theta_0}\frac{(1+r_0\cos2(\gamma-\phi_0)) e^{2\ii \lambda_{\parallel}} +(1-r_0\cos2(\gamma-\phi_0))  e^{2\ii \lambda_{\bot}}}{\sqrt{\left(1+ r^2_0\cos^22(\gamma-\phi_0) \right) -\left(1- r^2_0\cos^22(\gamma-\phi_0) \right)\cos2\left(\lambda_{\parallel}-\lambda_{\bot}\right)}}
\end{split}
\end{equation}

\section{Infinitesimal evolution after interaction}
\label{infevol}
Within the context of the GLKS dynamical system it is interesting to examine the case of an infinitesimal variation of the density matrix of the light after interaction with the quantum polariser by allowing the factor  $\cos(\lambda_{\parallel} -  \lambda_{\bot})$ to vary with time as $\approx 1 - \delta(t)$, $\delta \ll1$. Then $\cos^2(\lambda_{\parallel} -  \lambda_{\bot})\approx 1- 2\delta$, and we obtain for  Eqs.\;\eqref{rpp} -- \eqref{bp}:
\begin{equation}
\label{rppd}    r^{\prime}  \approx r_0(1-\delta \sin^22(\gamma- \phi_0))\approx r_0\exp-\delta \sin^22(\gamma- \phi_0)\, ,   
\end{equation}
\begin{equation}
\label{apd}   a   \approx \cos2\phi_0 - \delta\sin2(\gamma-\phi_0) \sin2\gamma\, , 
\end{equation}
\begin{equation}
\label{bpd}   b   \approx \sin2\phi_0 + \delta\sin2(\gamma-\phi_0) \cos2\gamma\, , 
\end{equation}
\begin{equation}
\label{phipd}   \tan\phi^{\prime}   \approx \tan\phi_0 + \delta\frac{\sin2(\gamma-\phi_0)}{\sin2\phi_0\cos\phi_0}\left[(\sin(2\gamma-\phi_0) -\cos\phi_0\sin2(\gamma-\phi_0)\right]\, . 
\end{equation}
Eq.\;\eqref{phipd} can be rewritten as 
\begin{equation*}
\tan\phi^{\prime} -  \tan\phi_0  \approx \frac{\phi^{\prime}-\phi_0}{\cos^2\phi_0}\approx \delta\frac{\sin2(\gamma-\phi_0)}{\sin2\phi_0\cos\phi_0}\left[(\sin(2\gamma-\phi_0) -\cos\phi_0\sin2(\gamma-\phi_0)\right]\,, 
\end{equation*}
and so 
\begin{equation}
\label{phipd1}
\frac{\Delta \phi}{\delta} \approx \frac{\sin2(\gamma-\phi_0)}{2\sin\phi_0}\left[(\sin(2\gamma-\phi_0) -\cos\phi_0\sin2(\gamma-\phi_0)\right]\,.
\end{equation}
By comparing Eq.\;\eqref{rppd} and Eq.\;\eqref{phipd1}   with the GKLS equations Eq.\;\eqref{dynsys2} and  Eq.\;\eqref{dynsys1} respectively in the case $\alpha=0$, one can identify the parameters $\delta$, $\beta$ , and $\mathcal{E}$: 
\begin{equation}
\label{infindel1}
\delta(t)= t-t_0 = \Delta t\ll 1\, , \quad \beta= \sin^22(\gamma-\phi_0)\,.
\end{equation}
\begin{equation}
\label{infindel2}
\mathcal{E}= - \frac{\sin2(\gamma-\phi_0)}{2\sin\phi_0}\left[(\sin(2\gamma-\phi_0) -\cos\phi_0\sin2(\gamma-\phi_0)\right]\,. 
\end{equation}

\subsection*{Acknowledgments} 
We thank the financial support from the Brazilian scientific agencies Fundação de Amparo à Pesquisa do Estado do Rio de Janeiro (FAPERJ), Coordenação de Aperfeiçoamento de Pessoal de Nível Superior (CAPES) and Conselho Nacional de Desenvolvimento Científico e Tecnológico (CNPq), grant related to Programa de Capacitação Institucional (PCI) 2018-2023, number 301098/2024-7. T.K. acknowledges the financial support by CNPq (No.\ 305654/2021-7) and the Fueck-Stiftung.
A part of this work has been done under the project INCT-Nuclear Physics and Applications (No.\ 464898/2014-5).

\end{document}